\documentclass[aps,prb, twocolumn, superscriptaddress]{revtex4}
\usepackage{graphicx}
\usepackage{amsmath}
\usepackage{amsfonts}
\usepackage{amsthm}
\usepackage{amssymb}
\usepackage{amsbsy}
\usepackage{pstricks}
\usepackage{wasysym}
\usepackage{mathrsfs}

\input{epsf}
\newcommand{\beq}{\begin{equation}}
\newcommand{\eeq}{\end{equation}}
\newcommand{\be}{\begin{eqnarray}}
\newcommand{\ee}{\end{eqnarray}}

\def\zhat{\hat{\mathbf{z}}}

\def\br{\mathbf{r}}
\def\bT{$\mathbf{T}$}

\begin{document}

\title{A Typology for Quantum Hall Liquids}
\author {S. A. Parameswaran}
\email{sidp@berkeley.edu}
\affiliation{Department of Physics, Princeton University, Princeton, NJ 08544, USA}
\affiliation{Department of Physics, University of California, Berkeley, CA 94720, USA}
\author {S. A. Kivelson}
\email{kivelson@stanford.edu}
\affiliation{Department of Physics, Stanford University, Stanford, CA 94305, USA}
\author {E. H. Rezayi}
\email{erezayi@calstatela.edu}
\affiliation{Department of Physics, California State University, Los Angeles, CA 90032, USA}
\author {S. H. Simon}
\email{s.simon1@physics.ox.ac.uk}
\affiliation{Rudolf Peierls Centre for Theoretical Physics, 1 Keble Road, Oxford University, OX1 3NP, UK}
\author {S. L. Sondhi}
\email{sondhi@princeton.edu}
\affiliation{Department of Physics, Princeton University, Princeton, NJ 08544, USA}
\author {B. Z. Spivak}
\email{spivak@u.washington.edu}
\affiliation{Department of Physics, University of Washington, Seattle, WA 98195, USA}
\date{\today}
\begin{abstract}

There is a close analogy between the response of a quantum Hall  liquid (QHL)
to a small change in the electron density and the response of a superconductor to an externally applied  
magnetic flux---
an analogy which is made concrete in the Chern-Simons Landau-Ginzburg (CSLG) formulation of the problem.
As the Types of superconductor are distinguished by this response, so too for QHLs: a typology can be introduced which is, however, richer than that in superconductors owing to the lack of any time-reversal symmetry relating positive and negative fluxes.
At the boundary between Type I and Type II behavior, the CSLG action has a ``Bogomol'nyi point,'' where the quasi-holes (vortices) are non-interacting -- at the microscopic level, this corresponds to the behavior of systems governed by a set of model Hamiltonians 
which have been constructed to render exact a large class of QHL wavefunctions.
 All Types of QHLs are capable of giving rise to quantized Hall plateaux.

\end{abstract}

\maketitle
\noindent

A transparent way in which to understand many properties of different quantum Hall phases is via the field theory of a two-dimensional charged superfluid coupled to a fictitious Chern-Simons (CS) gauge field \cite{ZHANG:1989p1367, Read:1989p1421, Girvin:1987p1415}.  A consequence of CS electrodynamics is that charges are bound to a fixed number of flux quanta. This equivalence of flux and charge  implies that the   condensed state -- which exhibits the Meissner effect, perfect conductivity and quantized vortex excitations -- corresponds to an incompressible phase with a quantized Hall conductance, whose quasiparticles carry fractional electric charge and statistics. While the original Chern-Simons-Landau-Ginzburg (CSLG) theory provides a convenient description of the Laughlin states and the Haldane-Halperin hierarchy \cite{Haldane1983:p1, Halperin1984:p1} as condensates of composite bosons \cite{ZHANG:1989p1367}, a parallel treatment in terms of composite fermions \cite{Lopez:1991p1} extends the CS approach to describe fractional quantum Hall (FQH) phases seen in the vicinity of even-denominator filling factors as condensates of fermion pairs \cite{MOORE:1991p58,GREITER:1992p1262}. Additional results, such as 
a global phase diagram in which the plateau transitions are related to an underlying superconductor to insulator transition\cite{Kivelson1992:p1}, can also be derived  within the CSLG formalism. There is thus a useful mapping between superconductivity and the FQHE.

Superconductors famously come in two varieties, which differ in their response to external magnetic fields: Type I superconductors phase separate into superconducting and normal regions, with flux concentrated in the latter, while Type II superconductors form an Abrikosov lattice of vortices, each carrying a single flux quantum. The analogy between superconductors and FQH phases suggests that there is a similar distinction between Type I and Type II QH liquids, manifested in dramatically different patterns of charge organization upon doping. While in the clean limit, Type II QH liquids exhibit  Wigner crystallization of fractionally charged excitations,  their Type I cousins would exhibit phase separation. This quite general dichotomy was pointed out only recently, when it was argued  that Type I behavior occurs in paired QH states when the pairing scale is weak \cite{Parameswaran2011:p1}.  While the focus of that work was the Pfaffian phase  in the vicinity of filling factor $\nu=5/2$, the results generalize implicitly to all paired states.

In this paper, we expand significantly on this work. First, we identify a class of ``Bogomol'nyi points'' of the CSLG theory which occur at the seperatrix between Type I and Type II behavior where
 the quasiholes are non-interacting even while
the charged excitation spectrum is gapped.  (Such field theories are also referred to as ``self-dual'' points, for reasons we will discuss below.) 
In the microscopic, lowest Landau level (LLL) formulation
they correspond to special Hamiltonians introduced with the purpose of rendering particular model
wavefunctions exact ground states. While some of what we say in both settings is not new, their
connection has not been discussed before. Second, we observe that such self-dual points can be
weakly perturbed to yield various Types of QHL. 
These include the traditional Type I and
II liquids, 
  frustrated Type I liquids, which exhibit short-distance phase separation frustrated by long-range repulsion \cite{Parameswaran2011:p1},
  and others which we will discuss below.
Especially striking are
Type I-II QHLs which exhibit Type I behavior for one sign of doping and Type II for the other --
this Type might generalize to time reversal(\bT)-breaking superconductors as well.
Note that all Types of QHL exhibit  quantized transport plateaux 
as a function of magnetic field or density, at least upon including weak disorder.

\noindent{\bf Chern-Simons Landau-Ginzburg Theory.} Near $\nu=1/k$, the description of QHLs as superconducting states of flux-charge composites is formalized in terms of a (bosonic) composite field $\phi$ that binds
$k$ flux quanta to an electron and interacts with a CS 
gauge field $(a_0,{\bf a})$ which
encodes the electron statistics, as captured by the (gauge-fixed)
Euclidean Lagrangian density
\be
\label{eq:ZHKham}
{\cal L}=
\bar\phi D_{\tau}\phi+\frac{\left|\mathbf{D}\phi\right|^2}{2m} + \lambda\left(|\phi|^2 - \rho\right)^2 +i a_0\frac{\nabla\times\mathbf{a}}{k\Phi_0}
\ee
where $D_0=\partial_{\tau} - ia_0$ and $D_{\mu} = \partial_{\mu} -i(a+A)_\mu $ are the covariant derivatives, the external field $B =\nabla \times \mathbf{A}$, the filling is $\nu \equiv {\rho \Phi_0 \over B}={1 \over k}$,  with $\Phi_0 =hc/e$ the  quantum of flux, and $a_0$ is a non-dynamical field which enforces the flux attachment constraint 
\be
b \equiv \nabla\times \mathbf{a} &= -k\Phi_0 |\phi|^2  \label{eq:CSdensity}
\ee
The static extrema of (\ref{eq:ZHKham}) satisfy (\ref{eq:CSdensity}) and the CSLG equations of motion
\begin{subequations}\label{eq:eoms}
\begin{align}
-\frac{\mathbf{D}^2\phi}{2m} + 2\lambda\left(|\phi(\br)|^2 -\rho\right)\phi &= 0\label{eq:NLS}\\
\mathbf{e} \equiv -\nabla a_0 &= k\Phi_0 \zhat\times\mathbf{j} \label{eq:CScurrent}
\end{align}
\end{subequations}
where $\mathbf{j} = \frac{1}{m} \text{Im}[\bar\phi\mathbf{D}\phi]$ is the current 
density.
The integral of ${\cal L}$ over space gives 
the energy of this state.
Precisely at $\nu = 1/k$---corresponding to the
pristine QHL---these are solved
by the uniform condensate configuration $\phi = \sqrt\rho e^{i\theta}$, $\mathbf{a} = -\mathbf{A}$
and $a_0=0$, which yields a state with zero energy.

\noindent{\bf The Bogomol'nyi Point:}  In search of vortex excitations we employ the Bogomol'nyi trick,  
$|\mathbf{D} \phi|^2 = |{D}_\pm \phi|^2 \mp  (b+B)|\phi|^2 \pm m\nabla\times\mathbf{j}$ (where  ${D}_\pm \equiv  D_x \pm i D_y$) to
express ${\cal L}$ in (\ref{eq:ZHKham}) for  any time-independent field configuration subject to the constraint  (\ref{eq:CSdensity}) as
\be\label{eq:Bogomolnyi}
{\cal L}=&&\frac{|{D}_\pm\phi|^2}{2m} 
+ \left(\lambda \pm\frac{k\Phi_0}{2m}\right)\left(|\phi|^2 - \rho\right)^2 
 \nonumber\\
&&\mp \frac{(B-B^*)|\phi|^2 +B^*(\rho-|\phi|^2)}{2m} 
\ee
where $B^*\equiv k\Phi_0\rho$.  Since the total charge is the integral over space of $|\phi|^2$, and is a conserved quantity, the term on the second line makes a configuration independent contribution to the energy.  Thus, the optimal vortex solutions are those that minimize the first two terms.
At the Bogomol'nyi point \footnote{The alternative choice of $\lambda=-k\Phi_0/2m$ is unstable.}, 
 $\lambda = \lambda_\text{B} \equiv k\Phi_0/2m$, the quartic term cancels for the lower sign in (\ref{eq:Bogomolnyi}),
so this reduces to finding solutions to the first order equation
\be\label{eq:SDE}
{D}_-\phi&\equiv \left(D_x -i D_y\right) \phi = 0 \label{eq:SDa}
\ee
along with the constraint (\ref{eq:CSdensity}) \cite{Dunne:1995book}.  Eq. \ref{eq:SDa} represents a form of self-duality as it can be rewritten as $D_a\phi =i\epsilon_{ab}D_b\phi$.

In (\ref{eq:ZHKham}), $\rho$ is formally related to the zero of energy which we are free to choose so that $\rho$  is equal to the average charge density. In this case, such a solution has an energy-density $\mathcal{E}= \frac{(B-B^*)\rho}{2m}$.
From Eq.
(\ref{eq:SDE}) 
it is straightforward to show that
$\mathbf{j} = \frac{1}{2m} \zhat\times\nabla |\phi|^2$.   Thus, the final extremal condition, Eq. \ref{eq:CScurrent}, is solved by 
$ a_0 = \lambda_\text{B} |\phi|^2 + {\rm const}$ where the constant is related to the chemical potential. 

The energy of the self-dual solutions of a specified vorticity is {\it independent} of the spatial distribution of that vorticity. As our signs correspond to vortices being quasiholes, it follows 
that such self-dual quasiholes do not interact at $\lambda = \lambda_\text{B}$.
The quasielectron (anti-vortex) solutions are {\it not} self-dual.  These interactions have not been studied in detail although it is
clear that they have a range of order the magnetic length, $\ell_B = (2\pi \Phi_0/B^*)^{1/2}$ and, from \cite{Tafelmayer1993:p1} that they are repulsive at very short distances and in their tails.
However, as their order parameter (density) profile is radially non-monotone, one cannot at present
rule out an intermediate regime of attraction. We will assume that that the quasielectrons at
the self-dual point repel at all distances, although almost nothing that we say will depend
on this assumption as we will clarify below. Finally, the gap to qh-qe
pairs is positive \cite{Tafelmayer1993:p1} and the quadratic mode
frequencies are all positive, whence the self-dual point is stable.

\noindent{\bf Perturbing the Bogomol'nyi Points:}  It follows from the above that at the Bogomol'nyi points, QHLs are Type II for
quasielectron doping and agnostic for quasihole doping. We now study the effect of small perturbations.
\begin{enumerate}
  \item If we change decrease/increase $\lambda$ by a small amount $\delta \lambda$, the quasiholes now experience an attractive/repulsive interaction resulting in a Type I/Type II QHL for hole doping. The quasielectron interaction will not change sign for sufficiently small $\delta \lambda$ of either sign. The result is a symmetric Type II QHL for $\delta \lambda > 0$ and an asymmetric Type I-II QHL for $\delta \lambda <0$ (i.e., Type I for quasihole and Type II for quasielectron doping\footnote{If our assumption that the quasi-electrons uniformly repel at $\lambda=\lambda_B$ turns out to be incorrect, the only effect is that Type I-II behavior would be seen for $\delta \lambda > 0$.}). 
  \item While the restriction to a local scalar self-interaction is natural in a superconductor, a more general non-local density-density interaction
      $
      \delta \mathcal{L} = \frac{1}{2}\int d\br' (|\phi(\br)|^2 -\rho) v(\br -\br') (|\phi(\br')|^2-\rho)
      $ 
      is  natural in the context of the Hall effect.  Perturbing the Bogomol'nyi pont with a long-range repulsive interaction (e.g. Coulomb) clearly results in a symmetric Type II QHL.
      
 \item      Let us  perturb the self-dual model with a term of this form  in which $v(\br)$ is a) attractive, b) weak enough to not close the gap to making a quasielectron-quasihole pair and c) has a range $\gg \ell_B$. Now the quasiholes {\it and} quasielectrons attract at long distances which is sufficient for both to phase separate at finite densities and thus to exhibit (macroscopic) Type I behavior for both signs of doping. However, for quasielectron doping, this macroscopic Type I behavior hides the competition between the short ranged repulsion at the self-dual point and the longer ranged attraction.
 If the attraction is sufficiently weak, then within the quasi-particle rich region, the quasiparticles still form a Wigner crystal, a behavior analogous to what has been called Type 1.5 in the superconducting context in \cite{Babaev:2005p1}.
 \item More generally, perturbing with additional, non-monotone interactions can generate various forms of charge order upon doping.
\end{enumerate}

\noindent{\bf Analogy with Superconductors:}
As noted previously, in superconductors the Bogomol'nyi point marks the boundary
between Type I and II behavior. The key difference from the QH case is that {\it both} 
$D_\pm$ can
be used to obtain vortex {\it and} anti-vortex solutions which are, naturally, related to each other simply by time reversal conjugation. Thus near  the 
Bogomol'nyi point and, indeed more generally,
\bT-invariant superconductors exhibit a symmetric response to flux doping. This raises the interesting question of whether \bT-breaking superconductors can exhibit asymmetric flux
doping - for instance, Type I-II behavior. Conversely, we note that weakly coupled paired QH states  
give rise to a LG theory of essentially the superconducting form \cite{Parameswaran2011:p1} where the symmetry in doping can
be traced to the particle-hole symmetry about the Fermi surface of
the parent composite Fermi liquid. In this limit
the paired states exhibit two different length scales---the pairing (coherence) length and the screening length (penetration depth), and thus exhibits symmetric frustrated Type I behavior  
at  weak coupling  where the coherence length  greatly exceeds the penetration depth. 

\noindent{\bf Microscopic Models:} We now turn to a large class of microscopic models for FQH states which realize the key properties of the Bogomol'nyi point of the CSLG theory and thus can
be perturbed to yield Type I QH fluids in exactly the same fashion. These are not new models---they have been constructed historically to render various desirable wavefunctions exact ground states, starting with the work of \cite{Trugman1985:p1,Haldane1983:p1}. However, the connection
of these models to Bogomol'nyi points in the CSLG theory has not been made before to our
knowledge.

\begin{figure}
\includegraphics{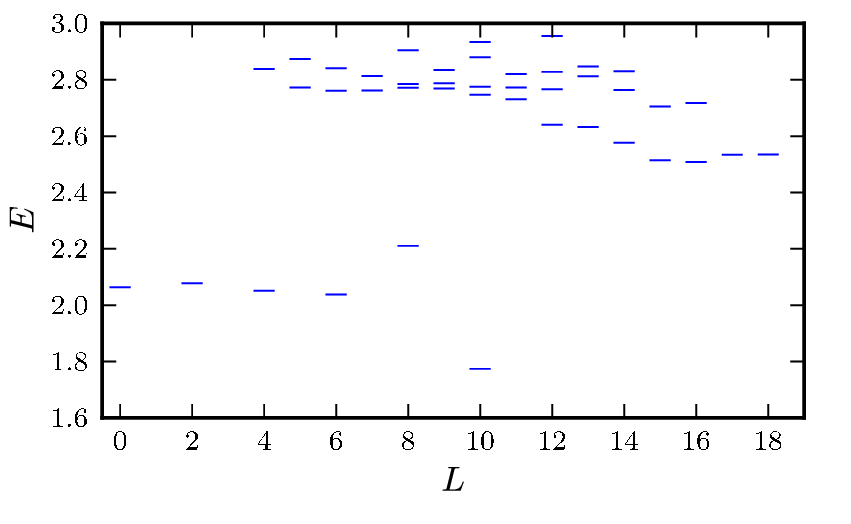}
\caption{\label{fig:qpspectrum} Eigenvalues of $H_{1/3}$ for $N=12$ electrons on a sphere and  flux corresponding to $\nu=1/3 +\text{two  quasielectrons}$. Two-quasielectron states have total angular momentum $L$ ($0\leq L\leq 12$), upon which their separation depends inversely. The `hard-core'  $L=12$ energy merges with the continuum. }
\end{figure}

An illustrative example of how this works is the $\nu=1/3$ state, in many ways the prototypical FQH state.  At this filling the ideal Laughlin state is the (essentially)
unique \footnote{The caveat allows for topological degeneracies on closed manifolds, in this case $3^g$ for genus $g$.} ground state of the model ``hard core'' or ``pseudo-potential''
Hamiltonian \cite{Trugman1985:p1,Haldane1983:p1}
$H_{1/3} = \sum_{i<j} \nabla^2 \delta^{(2)}(\br_i -\br_j)$ .
All available evidence is consistent with the proposition that at exactly $\nu=1/3$ the ground state is separated from 
all excited states 
by a gap
that remains non-zero in the thermodynamic limit. Further, {\it all} states with a given number of quasiholes are degenerate, or in other words,  the quasiholes do not interact. Quasielectrons on the other hand, do interact, although their interaction has not been analytically computed. {We have numerically evaluated the energy spectrum of  a system of 12 electrons on a sphere, at a flux density corresponding to two quasi-electrons --  the results  are shown in Fig.~\ref{fig:qpspectrum}.  Looking at the lowest energy states as a function of total angular momentum - which is inversely proportional to the distance between quasielectrons -- we see that they exhibit a short-ranged interaction consistent with a hard core, an intermediate attraction and an asymptotic
repulsion. We note that Beran and Morf\cite{Beran:1991p1} have also studied the quasiparticle interactions in the $1/3$ state for pseudopotentials tuned near the Coulomb point, which also exhibits non-monotonic features.} 
These features of the model Hamiltonian clearly parallel those of the Bogomol'nyi point of the CSLG theory; it follows that the model Hamiltonian is on the border between Type I and II
for hole doping and is Type II 
Type 1.5  for electron doping.
Therefore, we may follow the strategy adopted previously: by perturbing about $H_{1/3}$ with a weak, longer ranged interaction, we can make the quasiparticles either attract or repel without closing the gap and destabilizing the ground state. We have thus replicated our perturbative construction of various types of QH liquids at $\nu=1/3$ within a LLL treatment.

The $1/3$ Laughlin state is just one example of a much larger (indeed, infinite) class of  microscopic wavefunctions inspired by conformal field theory which are gapped and exact ground states of short-range Hamiltonians. It is believed that these criteria are satisfied by all states belonging to the so-called Read-Rezayi (RR) sequence \cite{ReadRezayi1999:p1}, and their particle-hole conjugates. The RR states have filling $\nu_{k,m} = k/(km+2)$, where $k$ is a nonzero positive integer, and $m$ is odd for QH states of fermions; their wavefunctions obey a generalized Pauli principle, and they can be obtained as the densest zero energy states of $k+1$ body model Hamiltonians.
 The $k=1$ case corresponds to the Laughlin states, while $k=2, m=1$ corresponds to the Moore-Read (Pfaffian) state \footnote{P-H conjugated RR states at  $\bar{\nu}_{k,m} =1-\nu_{k,m} $ reverse  the roles of quasiholes and quasielectrons and are ground states of conjugated Hamiltonians.}.

All these model Hamiltonians have noninteracting quasiholes and weakly (dominantly repulsively) interacting quasielectrons and thus exhibit the characteristics of a Bogomol'nyi point. It follows  
that by
perturbing them we can find various members of our doping typology. 

\noindent{\bf Plateau formation:} Plateau formation in a QHL refers to the invariance of the $T\to 0$
conductivity tensor as the density is varied (``doped'') over a non-zero range about the commensurate density, $\rho^*=B/k\Phi_0$.  For this to occur, the doped charge must be pinned, so it does not contribute to the DC transport.
For macroscopically Type II fluids this localization can arise from
disorder as commonly assumed in the theory of the QHE but potentially also from interactions
alone \footnote{Doped charges can be localized without disorder by forming quasiparticle/bubble crystals.}. For  
Type I fluids
with no disorder, the transport properties of the macroscopically phase separated state can depend on details of geometry and the nature of the boundary conditions.
However in two dimensions arbitrarily weak disorder prevents macroscopic phase separation \cite{imrywortis}
and leads to plateau formation as  discussed on phenomenological grounds in \cite{Kivelson1986:p1}. We note that as the charge of the minimum deconfined charged excitation is the same for all fluids derived from the same parent state, all of them will exhibit a combination
of activated and variable range hopping transport at low temperatures.

\noindent{\bf Experimental Realizations:} Thus far, we have been primarily concerned with a point of principle---establishing a doping typology of QH fluids.  To this end, 
we have
mostly 
considered model interactions
which differ substantially from those in typical experimental systems. We now comment briefly on the prospects for experimental realizations of the new members of this
typology: 
1) Apart from potential cold atom realizations, experimental systems involve repulsive Coulomb ($1/r$) or [in the presence of a nearby conducting plane] dipolar ($1/r^3$) interactions which limits the 
likely types to Type II, or frustrated Type I (which are thus macroscopically Type II)
QHLs. 
2) Paired QHLs currently appear to be the most promising 
candidates for states which exhibit quasiparticle clumping, as
they generically exhibit (frustrated) Type I behavior at weak pairing  \cite{Parameswaran2011:p1,Inprep}.

The prototypical paired state, the Pfaffian, can be tuned to weak coupling either by changing quantum well thickness \cite{Papic2009:p1}, or using graphene samples with a screening plane \cite{Papic2011:p1}. Another example is a bilayer system, with each layer at $\nu=1/2$. For large  layer separation $d\gg \ell_B$, the ground state is simply two decoupled composite Fermi liquids and hence compressible, but pairing of composite fermions between layers becomes increasingly favored as $d$ is decreased;  for $d\rightarrow 0$ the ground state is an interlayer paired  QH liquid \cite{MollerSimon2009:p1}. At intermediate $d\gtrsim\ell_B$, the pairing gap will be small reflecting weak coupling so the resulting paired state must be Type I. 
This is an example of a paired state that does not have a model microscopic Hamiltonian but nevertheless shows Type I behavior in the appropriate limit.

\noindent{\bf Concluding Remarks.}
In this paper, we have established a typology for doped QH liquids by perturbing about special
points where the quasiholes 
are non-interacting/weakly interacting and yet the
QH state is protected by a gap. Conversely, we have identified these special points as poised on the boundary of Type I/II behavior. For the CSLG theory this is tied to the mathematics of
self-duality. It is interesting to ask whether the self-dual
equations have meaning for model Hamiltonians such as $H_{1/3}$. 
It may be useful to note that at the self-dual point, $\mathbf{j}$ is the purely diamagnetic LLL current which {\it does} correctly
yield the current in quasihole states, and that the band mass $m$ drops out of the remaining
equations consistent with purely LLL physics \cite{SLSThesis}.

\noindent{\bf Acknowledgements.} We are grateful to D. Abanin, B. Estienne, Z. Papic, and R. Thomale for useful discussions.  This work was supported in part by NSF grants DMR-1006608 and PHY-1005429 (SAP, SLS), DMR-0758356 (SAK), and DMR-0704151 (BZS), and DOE grant {DE-SC0002140} (EHR).

\bibliography{TypeIQH_bib}
\end{document}